\documentstyle[aps,preprint]{revtex}
\begin{document}
\draft
\title{Flavor Condensate and Vacuum (In-)Stability in QED$_{2+1}$}
\author{Walter Dittrich and Holger Gies\thanks{E-mail address: 
holger.gies@uni-tuebingen.de}}

\address{Institut f\"ur theoretische Physik\\
          Universit\"at T\"ubingen\\
      Auf der Morgenstelle 14, 72076 T\"ubingen, Germany}
\date{}
\maketitle
\begin{abstract}
In this letter we study (2+1)--dimensional QED. The first part contains 
the computation of the flavor symmetry--breaking condensate and its 
relation to the trace of the energy--momentum tensor, while the second 
part is concerned with the computation of the effective action
allowing for non-constant static external magnetic fields. We do not
find that space derivatives in the magnetic field lower the energy of
the ground state as compared to a constant field configuration.
\end{abstract}
\pacs{PACS: 11.10.E, 12.20.}
\narrowtext
\section{Introduction}
During the last few years flavor symmetry--breaking in
(2+1)--dimensional QED has been at the center of interest for many
authors. For example, Gusynin et al. \cite{1} have demonstrated that
uniform magnetic background fields act as a catalyst for dynamical
flavor symmetry--breaking in 2+1 dimensions. Others \cite{2} have
looked at the impact of static inhomogeneous magnetic field
configurations on the ground state of QED$_{2+1}$. 

The problems attacked in the present work are quite similar; the
methods employed, however, and the results obtained are somewhat
different. In particular the results in the last chapter are in
contradiction to the findings of previous works. We also want to work
out some similarities and differences between QED$_{2+1}$ and the real
world of QED$_{3+1}$. This is done in section II. In section III we are
mainly concerned with the QED$_{2+1}$ vacuum structure and its
(in-)stability with respect to non-constant magnetic fields. Our
results are then summarized in section IV.

\section{Fermi--condensate and trace of the energy--momentum tensor in
QED$_{3+1}$ and QED$_{2+1}$}

Consider the Lagrangian of QED$_{3+1}$:
\begin{equation}
{\cal L}=-\bar{\psi} \left[ m+\gamma \left( \case{1}{\text{i}}
\partial -eA\right) \right]\psi\, . \label{1}
\end{equation}
When we take the derivative of the generating functional,

\begin{eqnarray}
Z=\text{e}^{\text{i}W}&=&\int {\cal D}\psi {\cal D}\bar{\psi} {\cal D}A\,
\text{e}^{-\text{i}\int \text{d}^4 x\, \bar{\psi}\left[m+\gamma \left(
\case{1}{\text{i}} \partial -eA\right)\right] \psi}\, ,\nonumber
\end{eqnarray}
i.e.,

\begin{eqnarray}
\frac{\partial \ln Z}{\partial m}&=&\frac{1}{Z} \frac{\partial
Z}{\partial m} \nonumber\\
&=&\frac{-\text{i}}{Z} \int {\cal D}\psi {\cal D}\bar{\psi} {\cal D}
A\,\left( \int \!\text{d}^4 x\, \bar{\psi}(x) \psi(x) \right)
\text{e}^{-\text{i} \int \bar{\psi} [\%] \psi} \nonumber\\
&=&-\text{i} \int \text{d}^4 x\, \langle \bar{\psi}(x) \psi(x) \rangle
\nonumber
\end{eqnarray}
or with $\ln Z=\text{i}W$ : $-\frac{\partial W}{\partial m}= \int
\text{d}^4 x \langle \bar{\psi}(x) \psi(x) \rangle$ , we obtain

\begin{equation}
\langle \bar{\psi} \psi \rangle =-\frac{\partial {\cal
L}^{(1)}}{\partial m} \, . \label{2}
\end{equation}
Here we introduced the one--loop effective Lagrangian ${\cal L}^{(1)}$
via 

\begin{displaymath}
W^{(1)}=\int \text{d}^4 x\, {\cal L}^{(1)} (x)\, .
\end{displaymath}
When we employ the Green's function

\begin{displaymath}
G(x,x'|A)=\text{i} \langle T(\psi(x)\bar{\psi}(x')) \rangle\, ,
\end{displaymath}
we can produce the following useful equalities:

\begin{equation}
\frac{\partial {\cal L}^{(1)}(x)}{\partial m} =-\text{i\, tr}\,
G(x,x|A) =-\langle \bar{\psi}(x)\,\psi(x)\rangle\, .\label{3}
\end{equation}
A variant of the above procedure implies for the trace of the
energy--momentum tensor

\begin{eqnarray}
\langle \Theta_\mu\,^\mu(x)\rangle &=&-\text{i}\, m\, \text{tr}\,
G(x,x|A)= m\frac{\partial {\cal L}^{(1)}(x)}{\partial m} \nonumber\\
&=&-m
\langle\bar{\psi}(x) \psi(x) \rangle\, . \label{4}
\end{eqnarray}
With the aid of Schwinger's proper--time technique \cite{3} we find
for a constant magnetic field configuration, $B=B_3=\text{const.}$ :

\begin{equation}
\langle \Theta_\mu\,^\mu\rangle(B)= \frac{eBm^2}{4\pi^2}
\int\limits_0^\infty \frac{\text{d}z}{z^2} \text{e}^{-\frac{m^2}{eB}
 z} \left[z\coth z-1-\case{1}{3} z^2 \right]\, .\label{5}
\end{equation}
After performing the integration, we get \cite{4}

\begin{eqnarray}
\langle \Theta_\mu\,^\mu\rangle(B)=&-&\frac{e^2B^2}{12\pi^2}
-\frac{m^4}{4\pi^2} \ln \frac{m^2}{2eB}
+\frac{eBm^2}{4\pi^2} \ln \frac{m^2}{2eB}  \nonumber\\
&+&\frac{m^4}{4\pi^2} +\frac{eBm^2}{2\pi^2}\left[ \ln\Gamma
\left(\case{m^2}{2eB} \right)- \case{1}{2} \ln 2\pi \right]\, .
\nonumber\\
\label{6}
\end{eqnarray}
We can also identify

\begin{eqnarray}
\langle\bar{\psi}\psi\rangle(B)&=&\frac{m^3}{4\pi^2} \ln \frac{m^2}{2eB}
-\frac{eBm}{4\pi^2} \ln \frac{m^2}{2eB} -\frac{m^3}{4\pi^2}
\nonumber\\
&&-\frac{eBm}{2\pi^2}\left[ \ln\Gamma
\left(\case{m^2}{2eB} \right)- \case{1}{2} \ln 2\pi \right]\,
.\label{7}
\end{eqnarray}
As a consistency check with Schwinger's formula for pair production
\cite{3}, we obtain

\begin{equation}
\text{Im}\langle\Theta_\mu\,^\mu\rangle(E) =m\frac{\partial}{\partial
m} \text{Im}{\cal L}^{(1)}(E) =-\frac{m^2eE}{4\pi^2} \sum_{n=1}^\infty
\frac{\text{e}^{-\frac{\pi m^2}{eE}n}}{n} .\label{8}
\end{equation}
Now we turn to a parity--invariant model of (2+1)--dimensional QED
with two flavors ($\pm$). The dynamics is contained in the Lagrangian
\cite{5}

\begin{eqnarray}
-\bar{\psi}\bigl[ m&+&\gamma\left(\case{1}{\text{i}} \partial
-eA\right)\bigr]\psi ={\cal L}= \sum_{\lambda=\pm} {\cal L}_\lambda
\nonumber\\
&=&-\sum_{\lambda=\pm} \bar{\psi}_\lambda \left[ \lambda
m+\gamma\left( \case{1}{\text{i}} \partial -eA \right) \right]
\psi_\lambda\, . \label{9}
\end{eqnarray}
On the left-hand side $\psi$ is a four--component spinor and the
$\gamma$'s are $4\times 4$ matrices whereas the $\psi_\pm$ are
two--component spinors and the $\gamma$'s on the right-hand side are
$2\times 2$ (Pauli--)matrices. We are working in 3--space with
$x=(x^0,x^1,x^2)$ and $g=\text{diag}\,(-1,1,1)$\, .

The fermion condensate can be written as

\begin{eqnarray}
W&&(x;m)\equiv \langle \bar{\psi}(x)\psi(x)\rangle =\sum_{\lambda=\pm}
\langle \bar{\psi}_\lambda(x) \psi_\lambda(x) \rangle_\lambda
\nonumber\\
&&=\sum_{\lambda=\pm} \lambda\, W_\lambda(x;m)\, ,\, W_-(x;m)
=W_+(x;-m) \nonumber\\
&&=\sum_{\lambda=\pm} \lambda\, W_+(x;\lambda m)=W_+(x;m)-W_+(x;-m)
\, ,\label{10}\\
W&&(x;-m)=-W(x;m)\, . \nonumber
\end{eqnarray}
We are interested in the condensate described by the parity--invariant
Lagrangian (\ref{9}):

\begin{eqnarray}
\langle\bar{\psi}(x)\psi(x)\rangle&=&\langle \psi^\dagger(x)\gamma^0_{
4\times 4} \psi(x)\rangle \quad ,\nonumber\\
\gamma^0_{4\times 4}& =&\left(
\begin{array}{cc} \gamma^0_{2\times 2}&0\\ 0& -\gamma^0_{2\times 2}
\end{array}\right)\, .\label{11}
\end{eqnarray}
Changing the sign of $\gamma^0_{2\times 2}=\sigma_3$ corresponds to
changing the sign of m. For the moment we choose $m>0$; but there will
be a $\text{sgn}(m)$ factor later on.

Earlier in this section we derived the relation

\begin{eqnarray}
\langle\bar{\psi}(x)&&\psi(x)\rangle =\text{i}\, \text{tr}\, G(x,x|A)
=\text{i}\,\text{tr}\int \frac{\text{d}^3 k}{(2\pi)^3} {\cal G}(k)
\nonumber\\
&&=-\frac{1}{(2\pi)^3} \int \text{d}^3k\int\limits_0^\infty
\text{d}s\,\text{e}^{\left\{-\text{i}s\left[m^2-k^{0^2} +\frac{\tan z}{z}
k^2_\perp\right] \right\}} \nonumber\\
&&\qquad\times \text{tr}\left[ \frac{\text{e}^{\text{i}\sigma_3 z}}{\cos
z} \left( m+\gamma^0 k^0 -\frac{\text{e}^{-\text{i}\sigma_3 z}}{\cos
z} (\gamma\cdot k)_\perp\right)\right] \nonumber\\
&&\qquad\qquad z=eBs \, . \label{12}
\end{eqnarray}
Here we employed formula (2.47) of ref.\cite{6}. The momentum integral
and the traces can readily be done with the intermediate result

\begin{equation}
\langle\bar{\psi}\psi\rangle =-\int\limits_0^\infty
\frac{\text{d}s}{8(\pi s)^{\frac{3}{2}}} 4m(eBs)\cot(eBs)
\text{e}^{-\text{i} \left( m^2s+\case{\pi}{4} \right)}\, .\label{13}
\end{equation}
To have a convergent expression on the right-hand side we need one
subtraction which changes (\ref{13}) into $(s=-\text{i}t)$:

\begin{equation}
\langle\bar{\psi}\psi\rangle(B)=-\frac{m}{2\pi^{\frac{3}{2}}}
\int\limits_0^\infty \text{d}t\, \text{e}^{-m^2t} t^{-\frac{1}{2}}
\left[ eB \coth (eBt) -\frac{1}{t} \right] \, .\label{14}
\end{equation}
At this stage we need the value of the two integrals which after
dimensional regularisation yield

\begin{eqnarray}
I_1&=&\int\limits_0^\infty \text{d}t\, \text{e}^{-m^2t} t^{-\frac{1}{2}}
(eB) \coth (eBt) \nonumber\\
&=& \sqrt{eB} \sqrt{\pi} \left[ \sqrt{2} \, \zeta \left(\case{1}{2},
\case{m^2}{2eB} \right) -\frac{\sqrt{eB}}{m} \right]\, , \nonumber\\
I_2&=&\int\limits_0^\infty \text{d}t\, \text{e}^{-m^2t}
t^{-\frac{3}{2}} =2\sqrt{\pi} m\, . \nonumber
\end{eqnarray}
The use of these integrals produces

\begin{equation}
\langle \bar{\psi}\psi\rangle(B)=-\frac{1}{2\pi} \left[ m\sqrt{2eB}\,
\zeta\left( \case{1}{2}, 1+\case{m^2}{2eB} \right) +eB -2m^2 \right]\,
. \label{15}
\end{equation}
This result should be read side by side with formula (\ref{7}).

We observe that for $m>0$ and $(eB)>0$ we obtain

\begin{equation}
\lim_{m\to 0_+} \langle \bar{\psi}\psi\rangle (B) =-\frac{eB}{2\pi}
\label{16}
\end{equation}
and more generally

\begin{equation}
\lim_{m\to 0_+} \langle \bar{\psi}\psi\rangle (B)
=-\text{sgn}(m)\frac{eB}{2\pi} \, .\label{17}
\end{equation}
Note that in 3+1 dimensions we would have obtained
$\langle\bar{\psi}\psi\rangle \propto m\ln m =0$ as $m\to 0$, i.e.,
there is no spontaneous flavor symmetry breaking in QED$_{3+1}$.
Employing equation (\ref{4}) we can make contact with the trace of the
energy--momentum tensor in QED$_{2+1}$:

\begin{eqnarray}
\langle \Theta_\mu\, ^\mu\rangle&=& -m\langle\bar{\psi}\psi\rangle
\nonumber\\
&=&\frac{1}{2\pi} \left[ m^2 \sqrt{2eB}\,\zeta \left(\case{1}{2},1+
\case{m^2}{2eB} \right) +meB -2m^3 \right]\, .\nonumber\\
 \label{18}
\end{eqnarray}
Unlike the result in 4--dimensional QED \cite{4},

\begin{displaymath}
\lim_{m\to 0} \langle \Theta_\mu\, ^\mu \rangle_{(4)}=-\frac{1}
{12\pi^2}e^2B^2 =-\frac{2\alpha}{3\pi} \, \frac{1}{4} 
F_{\mu\nu}F^{\mu\nu}
\end{displaymath}
there is no trace-anomaly in 3--dimensional QED:

\begin{equation}
\lim_{m\to 0} \langle \Theta_\mu\,^\mu\rangle =0\, ,\label{19}
\end{equation}
as it should.

Incidentally, the formula for the pair production rate in
QED$_{2+1}$, which should be read together with expression (\ref{8}),
is given by

\begin{equation}
2\text{Im} {\cal L}^{(1)}(E)=\frac{1}{4\pi^2} (eE)^{\frac{3}{2}}
\sum_{n=1}^\infty \frac{\text{e}^{-\frac{\pi
m^2}{eE}n}}{n^{\frac{3}{2}}}\, .\label{20}
\end{equation}

\section{QED$_{2+1}$ for non-constant fields}
We now turn to a disussion of external inhomogeneous electromagnetic
fields interacting with massive fermions and begin by introducing the
Hamiltonian $D\!\!\!\!/^{\,2}$ which acts in a fictitious quantum
mechanical Hilbert space and whose coordinate representation is given
by

\begin{equation}
\langle x|H|y\rangle=(D\!\!\!\!/_x^{\,2}) \delta(x-y)\, ,\, 
D\!\!\!\!/_x=\gamma^\mu (\partial_x^\mu -\text{i} eA^\mu(x) )\, 
.\label{21}
\end{equation}
We are working in Euclidean space and the $\gamma$--matrices are the
above-mentioned $2\times 2$ matrices $\gamma_{2\times 2}$. The time 
developement in this 3--dimensional space is governed by the operator

\begin{equation}
K(t)=\text{e}^{-Ht}\, , \quad t>0\, . \label{22}
\end{equation}
The well-known relation between the kernel $K(t)$ and the Green's
function $G$ is stated in

\begin{eqnarray}
G&=&\int\limits_0^\infty\text{d}t\, \text{e}^{-m^2t} \, K(t)
=\int\limits_0^\infty \text{d}t\, \text{e}^{-(H+m^2)t} \nonumber\\
&=&\frac{1}{D\!\!\!\!/^{\,2}+m^2}\, .\label{23}
\end{eqnarray}
We will be interested in the matrix element

\begin{equation}
K(x,y;t)=\langle x,t|y,0\rangle=\langle x|\text{e}^{-Ht}|y\rangle
\label{24}
\end{equation}
which satisfies the diffusion equation

\begin{equation}
\left(\frac{\partial}{\partial t} +D\!\!\!\!/_x^{\,2}\right)\, 
K(x,y;t)=0\label{25}
\end{equation}
with $K(x,y;0)=\langle x|y\rangle=\delta(x-y)$.

Introducing $\Pi_\mu=p_\mu-eA_\mu$ and $p_\mu=-\text{i}\partial_\mu$ we
can rewrite the Hamiltonian in the form
$(\sigma_{\mu\nu}=\case{\text{i}}{2}[\gamma_\mu,\gamma_\nu])$

\begin{eqnarray}
H&=&D\!\!\!\!/^{\,2}=-(\gamma\Pi)^2=\Pi_\mu\Pi_\mu -\case{e}{2}
\sigma_{\mu\nu} F_{\mu\nu} \nonumber\\
\text{or}
\quad H&=&-\partial_\mu\partial_\mu +\text{i}e(\partial_\mu A_\mu)
+\text{i}eA_\mu\partial_\mu \nonumber\\
&&+e^2 A_\mu A_\mu -\case{e}{2}\sigma_{\mu\nu}F_{\mu\nu}\, .\nonumber
\end{eqnarray}
When this Hamiltonian is substituted into (\ref{25}) we find for the
diffusion equation in QED$_{3}$

\begin{eqnarray}
\left( -\partial^2 +2\text{i}eA_\mu\partial_\mu
+X+\frac{\partial}{\partial t} \right) \,K(x,y;t)&=&0\, ,\label{26}\\
\text{with}\quad X=\text{i}e\partial_\mu A_\mu +e^2A_\mu A_\mu
&-&\case{e}{2} \sigma_{\mu\nu} F_{\mu\nu}\, .\nonumber
\end{eqnarray}
For vanishing fields equation (\ref{26}) can easily be solved by

\begin{equation}
K_0(x,y;t)=\frac{1}{(4\pi t)^{\frac{3}{2}}}
\text{e}^{-\frac{(x-y)^2}{4t}}\, .\label{27}
\end{equation}
For non-vanishing fields we try to solve (\ref{26}) by the ansatz

\begin{equation}
K(x,y;t)=\frac{1}{(4\pi t)^{\frac{3}{2}}}
\text{e}^{-\frac{(x-y)^2}{4t}} \sum_{k=0}^\infty a_k(x,y) \,t^k\,
.\label{28}
\end{equation}
This expression, when inserted in (\ref{26}), yields the following
recursion relation for the coefficients $a_k$:

\begin{eqnarray}
k=0\,:&&\quad(x-y)_\mu\, D_\mu\, a_0(x,y)=0\nonumber\\
k\geq 0\,:&&\quad\left(D^2+\case{e}{2} \sigma F\right) a_k(x,y) 
\nonumber\\
&&\qquad=(k+1)a_{k+1}(x,y)\nonumber\\
&&\qquad\quad+(x-y)_\mu (\partial_\mu -\text{i}eA_\mu)
a_{k+1}(x,y)\, . \nonumber
\end{eqnarray}
These recursion relations are formally the same as the ones that one
encounters in QED$_{4}$ \cite{7}. Furthermore, from our experience
with constant field configurations we decompose $K(x,y;t)$ into a
gauge independent factor and the counter-gauge factor

\begin{displaymath}
\Phi(x,y)=\exp\left\{ \text{i}e\int\limits_y^x \text{d}\xi_\mu\,
A_\mu(\xi)\right\}\, ,
\end{displaymath}
which allows us to separate off the gauge dependence from the
coefficients $a_k$ by writing

\begin{displaymath}
a_k(x,y)=\Phi(x,y)\, f_k(x,y)\, .
\end{displaymath}
Now we take over the known results from QED$_{4}$ \cite{7} 
and so obtain $([f_k](x)=\lim_{y\to x} f_k(x,y))$

\begin{eqnarray}
[f_0](x)&=&1\quad ,\quad [f_1](x)=\case{e}{2}\sigma F\, ,\nonumber\\
{[}f_2](x)&=&\case{e^2}{12}F^2+\case{e}{12}\sigma F^{,\mu\mu}+
\case{1}{2}\left[\case{e}{2} \sigma F\right]^2 \, ,\label{29}\\
{[}f_3](x)&=&\case{1}{3!} \left[\case{e}{2} \sigma F\right]^3
+\case{e}{2} \sigma F\left[ \case{e}{12} \sigma F^{,\mu\mu}+
\case{e^2}{12} F^2\right] \nonumber\\
&&+\case{e^2}{48} \sigma F^{,\mu}\sigma F^{,\mu}
+\case{e}{120} \sigma F^{,\mu\mu\nu\nu}-\case{e^2}{30}
F^{\alpha\beta} F^{\alpha\beta,\mu\mu} \nonumber\\
&&-\case{e^2}{45} (F^{,\mu})^2
-\case{e^2}{180} F^{\alpha\mu,\mu} F^{\alpha\nu,\nu}\, ,\label{30}\\
\text{where}&&\sigma F=\sigma^{\alpha\beta} F^{\alpha\beta}\, ,\,
F^{2}=F^{\alpha\beta} F^{\beta\alpha}\, ,\nonumber\\
&&(F^{,\mu})^2=F^{\alpha\beta,\mu}F^{\alpha\beta,\mu} 
\text{\,and}\, F^{\alpha\beta,\lambda}=\frac{\partial}{\partial
x_\lambda} F^{\alpha\beta}\, . \nonumber
\end{eqnarray}
These $[f_k](x)$ contain contributions with and without space
derivatives. Hence, it is only natural to write

\begin{equation}
\sum_{k=0}^\infty [f_k]\, t^k= (1+b_1(x)\,t +b_2(x)\, t^2+\dots)
\sum_{k=0}^\infty [f_k^{\text{c}}](x)\, t^k\, .\label{31}
\end{equation}
The coefficients $[f_k^{\text{c}}]$ depend on non-constant fields in
the same manner as they would on constant fields. The $b_i(x)$
represent corrections for non-constant fields. According to
(\ref{29}) and (\ref{30}) the terms without derivatives are given by

\begin{eqnarray}
[f_1^{\text{c}}]&=&\case{e}{2} \sigma F\quad ,\quad
[f_2^{\text{c}}]=\case{e^2}{12} F^2+\case{1}{2} \left[ \case{e}{2}
\sigma F\right]^2\, ,\nonumber\\
{[}f_3^{\text{c}}]&=& \case{1}{3!} \left[ \case{e}{2} \sigma F\right]^3
+\case{e}{2} \sigma F \, \case{e^2}{12}F^2\, .\label{32}
\end{eqnarray}
Substituting the results of (\ref{32}),(\ref{29}) and (\ref{30})
into (\ref{31}) produces

\begin{eqnarray}
b_1(x)&=&0\quad ,\quad b_2(x)=\case{e}{12} \sigma F^{,\mu\mu}
\nonumber\\
b_3(x)&=&\case{e^2}{48}\sigma F^{,\mu}\, \sigma F^{,\mu}
+\case{e}{120} \sigma F^{,\mu\mu\nu\nu} \label{33}\\
&&-\case{e^2}{30} F^{\alpha\beta} F^{\alpha\beta,\mu\mu}
-\case{e^2}{45} (F^{,\mu})^2 -\case{e^2}{180} F^{\alpha\mu,\mu}
F^{\alpha\nu,\nu} \, .\nonumber
\end{eqnarray} 
The series in (\ref{31}) containing the form of the constant field
contributions can be summed up into the well-known result \cite{3}

\begin{eqnarray}
K^{\text{c}}(x,y;t)&=& \case{1}{(4\pi t)^{\frac{3}{2}}}
\Phi(x,y)\exp \left\{\case{e}{2} \sigma Ft\right\} \label{34}\\
&&\times \exp\left\{ -\case{1}{2}\, \text{tr}\, \ln \left(
\case{\sin(eFt)}{eFt} \right) \right\} \nonumber\\
&&\times  \exp\left\{-\case{1}{4} (x-y) eF\cot (eFt)(x-y) \right\}\,
 .  \nonumber
\end{eqnarray}
In the limiting case $x\to y$ we then obtain for non-constant fields:

\begin{eqnarray}
K(x,x;t)&=& \case{1}{(4\pi t)^{\frac{3}{2}}} \exp\left\{ \case{e}{2}
\sigma Ft -\case{1}{2}\, \text{tr} \ln \left(\case{\sin(eFt)}{eFt}
\right) \right\}\nonumber\\
&&\times \left[1+\case{e}{12} \sigma F^{,\mu\mu} t^2+\left(
\case{e^2}{48} \sigma F^{,\mu} \sigma F^{,\mu}\right.\right. 
\label{35}\\
&&\quad\, +\case{e}{120} \sigma
F^{,\mu\mu\nu\nu}-\case{e^2}{30} F^{\alpha\beta} F^{\alpha\beta,\mu\mu}
-\case{e^2}{45} (F^{,\mu})^2 \nonumber\\
&&\left.\left.\quad\, -\case{e^2}{180} F^{\alpha\nu,\mu}
F^{\alpha\nu,\nu}\right) t^3+\dots\right]\, . \nonumber
\end{eqnarray}
Finally we arrive at the (Euclidean) one--loop effective action for
non-constant field configurations ($F_\mu^\ast=\case{1}{2}
\epsilon_{\mu\nu\rho} F_{\nu\rho}$):
\begin{eqnarray}
W^{(1)}&=&\int \text{d}^3x\, {\cal L}^{(1)}(x)\nonumber\\
& =&\frac{1}{2}
\int\text{d}^3x \int\limits_0^\infty \frac{\text{d}t}{t}
\text{e}^{-m^2t} \text{Tr} [K(x,x;t)+ \text{c.t.}] \label{36}
\end{eqnarray}
where
\begin{eqnarray}
\text{Tr}&& [K(x,x;t)+ \text{c.t.}] \label{37}\\
&&=\case{1}{(4\pi t)^{\frac{3}{2}}}
\text{tr}_\gamma \Bigl\{ \left( eF^\ast t\coth(eF^\ast t) +\case{e}{2}
\sigma Ft\right) \nonumber\\
&&\,\times\left[ 1+\case{e}{12} \sigma F^{,\mu\mu} t^2+\left(
\case{e^2}{48} \sigma F^{,\mu} \sigma F^{,\mu} +\case{e}{120} \sigma
F^{,\mu\mu\nu\nu}\right.\right.\nonumber\\
&&\left.\left.\left.
\, -\case{e^2}{30} F^{\alpha\beta}
F^{\alpha\beta,\mu\mu} 
-\case{e^2}{45} (F^{,\mu})^2 -\case{e^2}{180} F^{\alpha\mu,\mu}
F^{\alpha\nu,\nu}\right) t^3\right]\right\}. \nonumber
\end{eqnarray}
Deviating from QED$_{4}$ we now need the following trace identities
valid for QED$_{3}$ only:

\begin{eqnarray}
\text{tr}_\gamma&& 1=2\, ,\,\text{tr}\, \sigma^{\mu\nu} =0\,
,\,\text{tr}\, \sigma^{\mu\nu} \sigma^{\lambda\sigma}
=2(\delta^{\mu\lambda} \delta^{\nu\sigma} -\delta^{\mu\sigma}
\delta^{\nu\lambda}) \, ,\nonumber\\
\text{tr}&&\sigma^{\mu\nu} \sigma^{\lambda\sigma} \sigma^{\rho\kappa}
=2\text{i} (\epsilon^{\mu\nu\rho}\epsilon^{\lambda\sigma\kappa}
-\epsilon^{\mu\nu\kappa}\epsilon^{\lambda\sigma\rho}) \, .\nonumber
\end{eqnarray}
Performing the various trace operations in (\ref{37}) we finally end
up with

\begin{eqnarray}
W^{(1)}&=&\frac{1}{16\pi^{\frac{3}{2}}}
\int\limits_0^\infty\frac{\text{d}t}{t^{\frac{5}{2}}} \text{e}^{-m^2t}
\int \text{d}^3x \biggl\{ 2(eF^\ast t\coth (eF^\ast t)-1) 
\nonumber\\
&&+\case{e^2}{15} eF^\ast t \coth eF^\ast t \left( \case{7}{12}
F^{\alpha\beta,\mu}F^{\alpha\beta,\mu}
-F^{\alpha\beta}F^{\alpha\beta,\mu\mu} \right. \nonumber\\
&&\left.-\case{1}{6} F^{\alpha\mu,\mu}F^{\alpha\nu,\nu}\right)t^3
+\case{e^2}{6} F^{\lambda\sigma}F^{\lambda\sigma,\mu\mu}t^3
\nonumber\\
&&+\case{e^2}{60} F^{\lambda\sigma} F^{\lambda\sigma,
\mu\mu\nu\nu} t^4\biggr\} \, .\label{38}
\end{eqnarray}
Note that all terms on the right-hand side yield finite contributions.

In the sequel we will mostly be interested in non-constant (static)
external magnetic fields:

\begin{equation}
F^{12}=B\quad ,\quad F^\ast =B>0\, . \label{39}
\end{equation}
Then the first term in the curly brackets of (\ref{38}) yields the
well-known form for the effective Lagrangian in QED$_{3}$ for
constant $B$ fields \cite{8}:

\begin{equation}
{\cal L}^{(1)}_{\text{eff}}=\frac{\sqrt{2}}{2\pi} (eB)^{\frac{3}{2}}
\, \zeta\left( -\case{1}{2}, \case{m^2}{2eB}\right)-\frac{meB}{4\pi}\,
.\label{40}
\end{equation}
The correction terms for non-constant $B$ fields that follow from
(\ref{38}) are contained in three contributions:

\begin{equation}
{\cal L}^{(1)}_{\text{A,eff}}=\frac{1}{16\pi^{\frac{3}{2}}}
\frac{e^2}{15} \, Q\, \int\limits_0^\infty \text{d}t\, \sqrt{t} (eBt)
\coth (eBt) \, \text{e}^{-m^2t} \, ,\label{41}
\end{equation}
where

\begin{eqnarray}
Q&=&\left(\case{7}{12} F^{\alpha\beta,\mu}F^{\alpha\beta,\mu}-
F^{\alpha\beta}F^{\alpha\beta,\mu\mu} -\case{1}{6}
F^{\alpha\mu,\mu}F^{\alpha\nu,\nu} \right)\nonumber\\
&=&\left(\case{7}{6} \partial^\mu B\partial^\mu B-B\partial^2 B-
\case{1}{6} \partial^j B\partial^j B\right)\, ,\, j=1,2\, .\label{42}
\end{eqnarray}
The value of the integral occurring in (\ref{41}) turns out to be

\begin{displaymath}
\frac{1}{(eB)^{\frac{3}{2}}} \frac{3\sqrt{\pi}}{4} \left[
\frac{1}{2\sqrt{2}} \zeta\left(\case{5}{2},\case{m^2}{2eB} \right)
-\left(\frac{eB}{m^2} \right)^{\frac{5}{2}} \right] \, .
\end{displaymath}
Hence we obtain

\begin{equation}
{\cal L}^{(1)}_{\text{A,eff}} =\frac{\sqrt{e}}{320\pi}
\frac{Q}{B^{\frac{3}{2}}} \left[\frac{1}{2\sqrt{2}} 
\zeta\left(\case{5}{2},1+\case{m^2}{2eB} \right)
+\left(\frac{eB}{m^2} \right)^{\frac{5}{2}} \right]\, .\label{43}
\end{equation}
The last two terms in (\ref{38}) contribute

\begin{eqnarray}
{\cal L}^{(1)}_{\text{B,eff}}&=&\frac{1}{16\pi^{\frac{3}{2}}}
\frac{e^2}{6} F^{\lambda\sigma}F^{\lambda\sigma,\mu\mu}
\int\limits_0^\infty \text{d}t\, \sqrt{t}\,\text{e}^{-m^2t}\nonumber\\
&=&\frac{1}{96\pi} \frac{e^2}{|m|^3} (B\partial^2 B)\, ,\label{44}\\
{\cal L}^{(1)}_{\text{C,eff}}&=&\frac{1}{16\pi^{\frac{3}{2}}}
\frac{e^2}{60} F^{\lambda\sigma}F^{\lambda\sigma,\mu\mu\nu\nu}
\int\limits_0^\infty \text{d}t\, t^{\frac{3}{2}} \text{e}^{-m^2t}
\nonumber\\ 
&=&\frac{1}{640\pi} \frac{e^2}{|m|^5} (B\partial^2 \partial^2B)\, 
.\label{45}
\end{eqnarray}
Finally we come to the limiting case of vanishing fermion masses
$(m\to 0)$. Let us regard the convergent terms of (\ref{40})
and (\ref{43}) first:

\begin{eqnarray}
{\cal L}^{(1)}_{\text{eff}}&=&-\frac{1}{4\sqrt{2}\pi^2}
\,(eB)^{\frac{3}{2}}\, \zeta\left(\case{3}{2}\right) \, ,\label{46}\\
{\cal L}^{(1)}_{\text{A,eff}}&=&\frac{1}{4\sqrt{2}\pi^2} 
\frac{\pi}{160}\,\frac{\sqrt{e}\,Q}{B^{\frac{3}{2}}}\, 
\zeta\left( \case{5}{2}\right)\, .\label{47}
\end{eqnarray}
Now, for static inhomogeneous $B$ fields we have

\begin{equation}
Q=\partial^j B\partial^j B-2B \partial^2 B \label{48}
\end{equation}
so that

\begin{equation}
-\frac{1}{B^{\frac{3}{2}}} 2B\partial^2 B=2\partial_\mu
\left(\frac{1}{B^{\frac{1}{2}}} \right) \partial^\mu B
=-\frac{\partial^jB \partial^jB}{B^{\frac{3}{2}}} \label{49}
\end{equation}
which exactly cancels the first term in (\ref{48}). Hence in the limit
$m\to 0$ there are no corrections to the one--loop effective
Lagrangian that arise from the inhomogeneity of the static $B$ field.

There remains the problem with the diverging $(m\to 0)$ contributions
contained in (\ref{43}),(\ref{44}) and (\ref{45}). These can be made
to vanish by formulating QED$_{2+1}$ in a parity--invariant manner by
adding another flavor degree of freedom as done in section 2. The
result is a model with four-component spinors $\psi$, which is
equivalent to a theory describing two species of two-component spinors
$\psi_\pm$, one with mass $+m$ and the other with mass $-m$. So far we
chose $m$ to be positive. In our new extended version of QED$_{2+1}$
we now encounter the mass term

\begin{displaymath}
-m\bar{\psi}\psi=-m\psi^\dagger\gamma^0_{4\times 4}\psi=-m\psi^\dagger_+
\gamma^0_{2\times 2}\psi_+ +m\psi^\dagger_- \gamma^0_{2\times 2} \psi_-
\end{displaymath}
\begin{displaymath}
\text{where we used}\, \gamma^\mu_{4\times
4}=\left(\begin{array}{cc} \gamma^\mu_{2\times 2}&0\\0&
-\gamma^\mu_{2\times 2} \end{array} \right)\, \text{and}\,
\gamma^0_{2\times 2} =\sigma_3 .
\end{displaymath}
The terms in the effective Lagrangian that remain finite (after taking
the limit $m\to 0$) carry the same sign. The others, however, that
vanish or are singular for $m\to 0$ have their origin in the mass term
of the original Lagrangian and thus contain different signs for the
two different two-component fermion fields. Since they have the same
modulus they cancel each other.

\section{conclusion}
In the first half of this paper we studied the Fermi condensate in
QED$_{2+1}$. We showed that the flavor condensate is non-zero as the
fermion mass goes to zero. This is a specific (2+1)--dimensional
phenomenon since there is no spontaneous symmetry breaking in ordinary
QED$_{3+1}$. We also compared the trace anomaly of the
energy--momentum tensor and the electron--positron pair production
rate in both theories.

In the second half we employed the heat kernel expansion to determine
the Seeley coefficients that are necessary to calculate the effective
action of QED$_{2+1}$. We investigated the QED$_{2+1}$ ground state by
allowing the static external magnetic field to become space dependent.
Contrary to the result of reference \cite{2} we find that the ground
state probed in one-loop approximation is not shifted towards a lower
value by the presence of an inhomogeneous static magnetic field -- for
$m\to 0$. Our calculations do not exhibit an instability of the
uniform magnetic field state towards a more disordered state with
inhomogeneous magnetic fields. Although we believe that the stable
QED$_{2+1}$ ground state tested with an external constant magnetic
field is only an intermediate step on the way to the true ground
state, we do not find -- to one--loop order -- that space
inhomogeneities in the magnetic field change the situation.

\end{document}